\documentclass[onecolumn,amsmath,amssymb,12pt,superscriptaddress,nofootinbib]{revtex4}
\pdfoutput=1

\usepackage[latin1]{inputenc}
\usepackage[english]{babel}
\usepackage{amssymb}
\usepackage{amsmath}
\usepackage{amsthm}
\usepackage[]{graphicx}
\usepackage[]{subfigure}
\usepackage{tensor}
\usepackage{color}
\usepackage{cancel}
\usepackage{setspace}
\usepackage{fancyhdr}
\usepackage{framed}
\usepackage{tikz}
\usepackage[bookmarks,linktocpage, colorlinks=true, plainpages = false, citecolor = blue,  linkcolor=blue, urlcolor = blue, filecolor = blue]{hyperref} 

\begin{document}

\allowdisplaybreaks
\begin{titlepage}

\title{Unstable no-boundary fluctuations \\ from sums over regular metrics \\ $^{}$}

\author{Alice Di Tucci}
\email{alice.di-tucci@aei.mpg.de}
\affiliation{Max--Planck--Institute for Gravitational Physics (Albert--Einstein--Institute), 14476 Potsdam, Germany}
\author{Jean-Luc Lehners}
\email{jlehners@aei.mpg.de}
\affiliation{Max--Planck--Institute for Gravitational Physics (Albert--Einstein--Institute), 14476 Potsdam, Germany}

\begin{abstract}
\vspace{.3in} \noindent It was recently shown by Feldbrugge et al. that the no-boundary proposal, defined via a Lorentzian path integral and in minisuperspace, leads to unstable fluctuations, in disagreement with early universe observations. In these calculations many off-shell geometries summed over in the path integral in fact contain singularities, and the question arose whether the instability might ultimately be caused by these off-shell singularities. We address this question here by considering a sum over purely regular geometries, by extending a calculation pioneered by Halliwell and Louko. We confirm that the fluctuations are unstable, even in this restricted context which, arguably, is closer in spirit to the original proposal of Hartle and Hawking. Elucidating the reasons for the instability of the no-boundary proposal will hopefully show how to overcome these difficulties, or pave the way to new theories of initial conditions for the universe.
\end{abstract}
\maketitle

\end{titlepage}

\tableofcontents

\section{Introduction}

It is an interesting question to what extent the history of our universe is characterised by inevitable events, and to what extent by accidental developments. It is clear that an understanding of the beginning of our universe, or, preferably phrased, an understanding of the emergence of spacetime and matter, would provide substantial clues in answering this question. Various possible answers have been considered, including: that there was no beginning (e.g. that spacetime is geodesically complete \cite{Bars:2011aa}), that the universe bootstraps itself \cite{Boyle:2018tzc}, that large quantum fluctuations create universes with random values for many parameters \cite{Garriga:1997ef} or that it re-creates itself more systematically \cite{Steinhardt:2001st,Lehners:2009eg}. Another option would be that spacetime arose from some more fundamental, non-geometric structures \cite{Banks:1996vh,Gielen:2013naa}. Here we will be concerned with the idea that the universe is finite and entirely self-contained in space and in time. In the path integral approach to (semi-classical) quantum gravity, this idea can be taken to mean that one should sum only over 4-geometries that have no boundary to the past -- hence the name ``no-boundary proposal'' \cite{Hawking:1981gb,Hartle:1983ai}. In a closely related manner, one may view this prescription as expressing the idea that the universe could have tunneled in a smooth manner from nothing \cite{Vilenkin:1982de}. 

The no-boundary proposal was inspired by Euclidean quantum gravity, since in Euclidean signature it is straightforward to find solutions to the gravitational field equations that indeed have no boundary, while in Lorentzian signature this is impossible. Of course, since the universe is Lorentzian at later times, one needs geometries that also contain a Lorentzian part, and in fact one is in general required to consider complex geometries, as we will discuss in more detail below. The question is then rather how the path integral should be fundamentally defined. In that regard, the Euclidean approach has always been plagued by the problem that the gravitational action is unbounded above and below (both because the kinetic term for the scale factor of the universe has the ``wrong'' sign, and also when allowing for non-trivial topologies \cite{Gibbons:1977zz}), so that it was never clear whether the Euclidean path integral was meaningful at all. A series of recent works has analysed this setting again, but from the point of view of the Lorentzian path integral. In this case, the integral is over phases $e^{iS/\hbar},$ where $S$ denotes the action for gravity coupled to a positive cosmological constant $\Lambda>0.$ This is a conditionally convergent integral, and care must be taken in defining it. Probably the most important result of \cite{Feldbrugge:2017kzv} was that the Lorentzian path integral can indeed be defined, at least in minisuperspace, and that it is unique. In these works, the mathematical framework of Picard-Lefschetz theory was used to re-write the Lorentzian path integral in such a way that it becomes absolutely convergent. (One may also prove the convergence without recourse to Picard-Lefschetz theory, see \cite{Feldbrugge:2018gin}). The same framework allows one to see rather clearly that the Euclidean path integral is divergent and consequently meaningless, at least when $\Lambda>0.$ 

When the Lorentzian path integral was evaluated imposing the no-boundary condition, it was however found that the fluctuations are unstable, in that small fluctuations obey an inverse Gaussian distribution \cite{Feldbrugge:2017fcc}. This implies that the no-boundary proposal leads to unphysical results, and must be abandoned as a possible initial condition for the universe. This negative result had as a consequence that other definitions of the gravitational path integral were tried out, using complex integration contours for the lapse integral \cite{DiazDorronsoro:2017hti,DiazDorronsoro:2018wro}. These mathematically, but not physically, motivated proposals were found to lead to inconsistencies in terms of physical interpretation \cite{Feldbrugge:2017mbc,Feldbrugge:2018gin}.  

Here we will pursue a different avenue: in the minisuperspace calculations just mentioned, the no-boundary condition is imposed as the condition that the universe should start out at zero size. Then it was found that the saddle points of the integral are all complex, and can be represented as half of a Euclidean 4-sphere glued onto half of a Lorentzian de Sitter hyperboloid. Thus the saddle point geometries are indeed regular, with the locus of zero scale factor simply being a regular point on a sphere. However, the minisuperspace integral itself contains many off-shell geometries that are singular where the universe has zero size. Could it be that the instability of the fluctuations is due to these singular off-shell geometries? Should one sum only over purely regular geometries?  On the one hand, it seems unlikely that the instability will disappear, since the relevant saddle point of the Lorentzian path integral is regular, yet also unstable. On the other hand, it seems worthwhile to investigate a restricted sum over purely regular geometries, since this seems to be more closely in the spirit of the original formulation of the no-boundary proposal \cite{Hawking:1981gb,Hartle:1983ai}. 

\begin{figure}
\begin{center}
\includegraphics[scale=0.5]{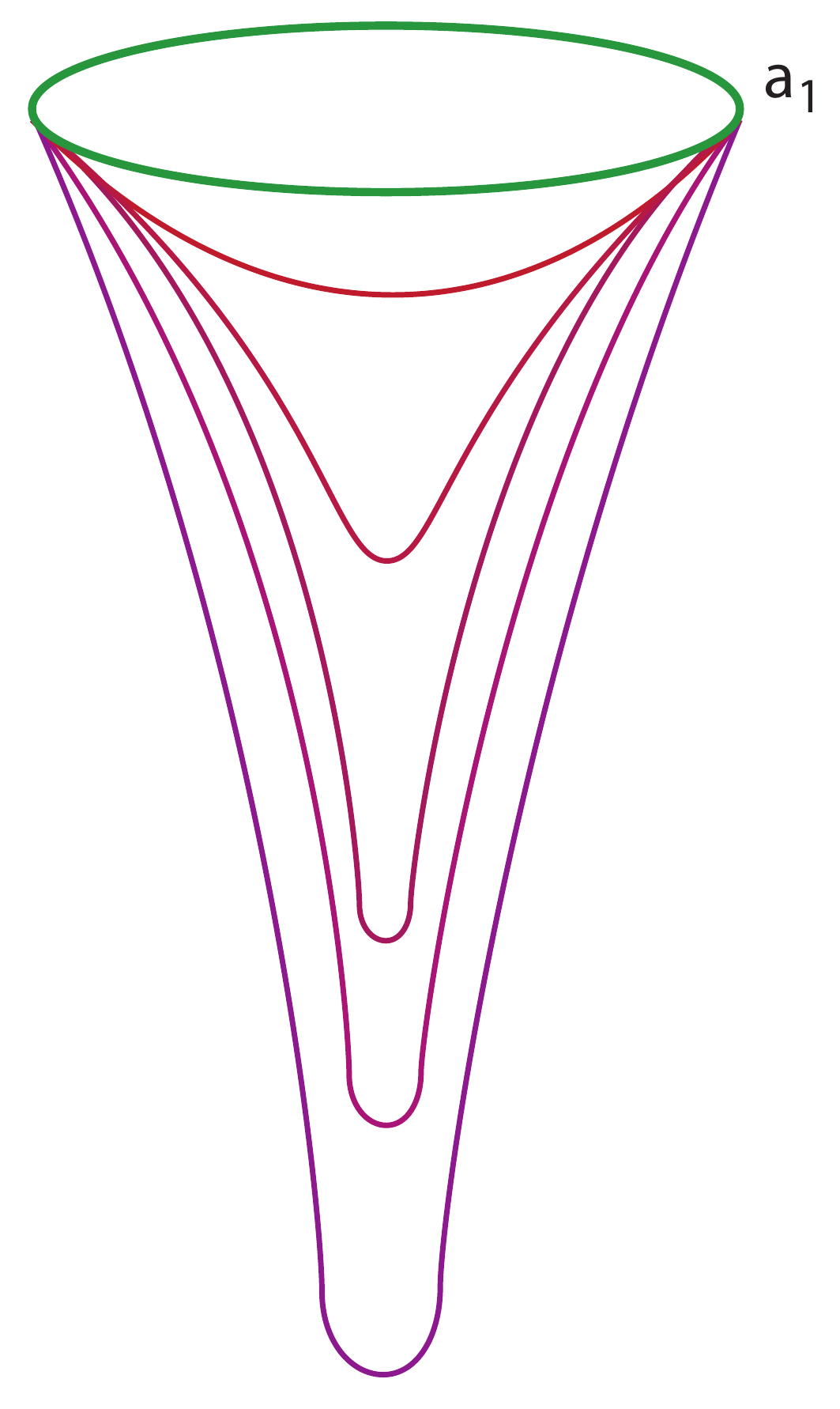}
\caption{We will consider a path integral over purely regular geometries, having as their only boundary a large late time universe with scale factor $a_1.$ The integral can be pictured as a sum over complexified 4-spheres, bearing in mind that a complexified 4-sphere contains a Lorentzian de Sitter section. One may argue that such a restricted sum is closest in spirit to the original idea of the no-boundary proposal \cite{Hawking:1981gb,Hartle:1983ai}.}
\label{fig:regular}
\end{center}
\end{figure}

Thus in the present paper we will consider a sum over regular geometries, as pictured in Fig. \ref{fig:regular}. Note that this is a highly restricted class of 4-geometries, as gravity has a tendency of causing collapse to a singularity. We will implement this sum over regular geometries by taking as our starting point a paper by Halliwell and Louko \cite{Halliwell:1989vu}, where they considered an integral over complexified 4-spheres of various radii. We will extend their work both for the background in section \ref{sec:bgd} (it turns out that they only considered half of all such 4-sphere solutions)  and then by adding perturbations in section \ref{sec:perts}. In the minisuperspace calculations of \cite{Feldbrugge:2017kzv}, the path integral contained separate integrals over the scale factor of the universe $r$ and over the lapse function $N$. The present setting is so restrictive that these two integrals get combined into a single ordinary integral over a variable $z$ which depends on both $r$ and $N$. One consequence of this is that it becomes impossible to define a purely Lorentzian integral, but we will discuss how one can define an integration contour that corresponds to the Lorentzian path integral as closely as possible, and we will also discuss other choices of contour. 

Going beyond a treatment of the background, we find that the action for the perturbations contains poles in the complex $z$ plane (where $z$ is the variable combining scale factor and lapse), with the location of the poles being wavenumber-dependent. This imposes additional restrictions on possible integration contours for the $z$ integral, as the contour must remain sensible and well defined for all possible choices of fluctuations. Here we will show that the analogue of the Lorentzian integral remains well defined after adding perturbations, but that a closed contour for instance would be adversely affected.

Our final result of the perturbative analysis is that the saddle points that were unstable in the minisuperspace setting remain unstable here, though now in a context where every summed over geometry is regular. Moreover, it is straightforward to verify in the present context under what conditions the backreaction of the perturbations remains small. Our conclusion will be that the instability of the no-boundary proposal is robust and in full agreement with the minisuperspace results.

\section{Background} \label{sec:bgd}

We are interested in evaluating the (suitably gauge fixed \cite{Teitelboim:1981ua,Teitelboim:1983fk}) path integral
\begin{align}
G[a_1;0]=\int^{a_1} {\cal D} g \, e^{iS/\hbar}
\end{align}
where $S=\frac{1}{2}\int d^4x \sqrt{-g}\left( R - 2 \Lambda \right)$ is the action for gravity plus a positive cosmological constant $\Lambda.$ The integral should be over metrics which have as their only boundary a final 3-surface with scale factor $a_1.$ A spatially homogeneous, isotropic and closed metric can be written as 
\begin{equation}
ds^2 = -N_L^2 dt^2 + a(t)^2 d \Omega_3^2
\end{equation}
where $d \Omega_3^2 $  is the unit 3-sphere with volume $2\pi^2$ and we will scale the time coordinate such that $0 \leq t \leq 1$. In light of the results of \cite{Feldbrugge:2017kzv,Feldbrugge:2017mbc}, we are interested in a sum over four metrics with \textit{Lorentzian} signature ($N_L$ real). However, in order to sum over regular (though complexified) four spheres we will use coordinates with \textit{Euclidean} signature, as this choice greatly facilitates the imposition of the no-boundary condition. Note that this is simply a coordinate choice (and since we will consider complexified metrics anyway this choice is truly arbitrary) -- the requirement that the path integral should be the Lorentzian one will then be implemented as a restriction on the possible integration contours. With the metric above, the action reads
\begin{equation}
S = 2 \pi^2 \int_0^1 N_L dt \left[ - 3 \frac{a \dot{a}^2}{N_L^2} - \Lambda a^3 + 3 a \right]\,,
\end{equation}
or, in terms of the Euclidean lapse $N_L = - i N_E$, 
\begin{equation}
S =2 \pi^2 i \int_0^1 N_E dt \left[ - 3 \frac{a \dot{a}^2}{N_E^2} + \Lambda a^3 - 3 a \right]\,.
\end{equation}
The  associated constraint then reads
\begin{equation}
3 \frac{\dot{a}^2}{a^2} - \frac{3 N_E^2}{a^2} + \Lambda N_E^2 = 0 
\end{equation}
and, with the boundary conditions $a(t=0)=0$ and $a(t=1)=a_1$ its solution is of the form 
\begin{equation}
a(t)= \pm \sqrt{\frac{3}{\Lambda}} \sin\left(\sqrt{\frac{\Lambda}{3}} N_E t\right) \label{asaddle}
\end{equation}
provided we choose $N_E$ such that $a(t = 1) = \pm \sqrt{\frac{3}{\Lambda}} \sin \left(\sqrt{\frac{\Lambda}{3}} N_E \right) = a_1 $. For both choices of sign, this metric then represents a 4-sphere of radius $\sqrt{3/\Lambda},$ which in general will be complexified since $N_E$ will take complex values whenever $a_1 > \sqrt{3/\Lambda}$. 

In the path integral we are not directly interested in the solutions to the equations of motion or the constraints, rather we first want to state the class of metrics that are to be summed over. Following \cite{Halliwell:1989vu} we will consider the simplest and most symmetric possibility, namely we will sum over 4-spheres with given  boundaries $a_0=0$ and $a_1 >0$ and arbitrary radius, 
\begin{equation}
a(t)= \pm r \sin \left(\frac{N_E \, t}{r}\right)  \label{eq:summedmetrics}
\end{equation}
with $a_1 = \pm r \sin \left( \frac{N_E}{r}\right)$ and $\dot{a}_1 = \pm N_E \cos \left( \frac{N_E}{r}\right)$. Accordingly, one should think of $N_E$ as being fixed by the boundary conditions and the sum to be over $r$. Given that $N_E$ will in general be a complex number, we should also expect $r$ to be complex, and that the integral will be over a contour in the complex $r$ plane. We note that in \cite{Halliwell:1989vu} only one choice of sign in \eqref{eq:summedmetrics} was considered. As we will see below, it is important to keep both signs at first, and then we will see how the sign may be fixed according to the integration contour chosen.

The action for the positive/negative choice of sign in \eqref{eq:summedmetrics} reads respectively
\begin{equation}
S = \pm \frac{i}{3 N_E } \left( 3 r^2 \left[ - 4 + 3 \cos \left(\frac{N_E}{r}\right)+ \cos^3 \left(\frac{N_E}{r}\right)\right] +  \Lambda r^4 \left[ 2 - 3 \cos \left(\frac{N_E}{r}\right)+ \cos^3 \left(\frac{N_E}{r}\right)\right]  \right) \label{eq2}
\end{equation}
It is possible to simplify the analysis by defining a new variable $z$ such that 
\begin{equation}
z = 1 + \cos \left( \frac{N_E}{r} \right) = 1 \pm \frac{\dot{a}_1}{N_E}\,, \label{z}
\end{equation}
implying the useful relation
\begin{align}
r^2 = \frac{a_1^2}{z(2 -z)}\,.
\end{align}
The action given by Eq. (\ref{eq2}) then reads in this variable
\begin{equation}
S = \pm 2 \pi^2 i a_1^2 \left[-z+ 1  + \left(\frac{a_1^2 \Lambda}{3}  - 4 \right)\frac{1}{z} + \frac{\Lambda}{3} \frac{a_1^2}{z^2} \right]\,,
\end{equation}
which diverges for $z \rightarrow 0 $ and $z \rightarrow \infty$.

Depending on the argument of $z $ the action diverges to $+ i \infty$ or $- i \infty$. This divides the complex $z$ plane into regions of convergence and divergence of the path integral $\int dz \, e^{ i S}$ (we will take the simplest measure in $z$, since we will only be interested in the leading terms in $\hbar$).  For $|z| \rightarrow \infty$, $ S \approx \mp i z \equiv \mp i R e^{i \theta}$, therefore the integrand $e^{ i S}$ diverges or vanishes in this limit depending on $Arg(z)$. For the choice $a = - r \sin ( \frac{N t}{r}),$ for instance, we have the limits
\begin{align}
\lim_{R \rightarrow \infty } e^{ R \cos \theta } e^{ i R \sin \theta} = \infty \; \; \; \; \; \; & \mbox{ for     }  \; \;  \frac{\pi }{2} < \theta < \frac{3 \pi }{2} \\
\lim_{R \rightarrow \infty } e^{ R \cos \theta } e^{ i R \sin \theta} = 0 \; \; \; \; & \mbox{ for } \; \;\;   - \frac{\pi }{2} < \theta < \frac{ \pi }{2}
\end{align}
Therefore, as $z$ goes to infinity along a direction exactly parallel to the imaginary line, the integrand $e^{iS}$ is purely oscillating. As soon as it slightly deviates from that direction, the path integral is either convergent or divergent. For $|z| \rightarrow 0,$ the action can be approximated by $S \approx \pm \frac{i}{z^2} = \pm \frac{i}{R^2} e^{ - 2 i\theta}.$ Thus the convergence regions in the small $z$ limit are the wedges $- \frac{\pi}{4} < \theta < \frac{\pi}{4}$ and $- \frac{3 \pi}{4} < \theta < \frac{5 \pi}{4}$ for $ a = + r \sin \left( \frac{N t}{r} \right)$ and respectively $\frac{\pi}{4}<  \theta < \frac{ 3 \pi}{4}$ and $\frac{5 \pi}{4}<  \theta < \frac{ 7 \pi}{4}$ for $a = - r \sin \left( \frac{N t}{r} \right)$.

The three saddle points for each sign of the action are located at 
\begin{align}
z_2 &= - 2 \\
z_{1,3} &= 1 \pm i \sqrt{\frac{a_1^2 \Lambda}{3} - 1} \,,
\end{align}
while the action at the saddle points is given by (where the signs are correlated with \eqref{eq:summedmetrics})
\begin{align}
S(z_2) &= \pm 2 \pi^2 i \, a_1^2 \left[ 5 - \frac{a_1^2 \Lambda}{12}\right]\,, \label{s1} \\
S(z_1) & = \mp \frac{ 12 \pi^2}{\Lambda} \left[i - \left(\frac{a_1^2 \Lambda}{3} -1 \right)^{3/2}\right]\,, \label{s2} \\
S(z_3) & = \mp \frac{ 12 \pi^2}{\Lambda} \left[i + \left(\frac{a_1^2 \Lambda}{3} - 1 \right)^{3/2}\right]\,. \label{s3}
\end{align}

\begin{figure}
\begin{center}
\includegraphics[width=0.4\linewidth]{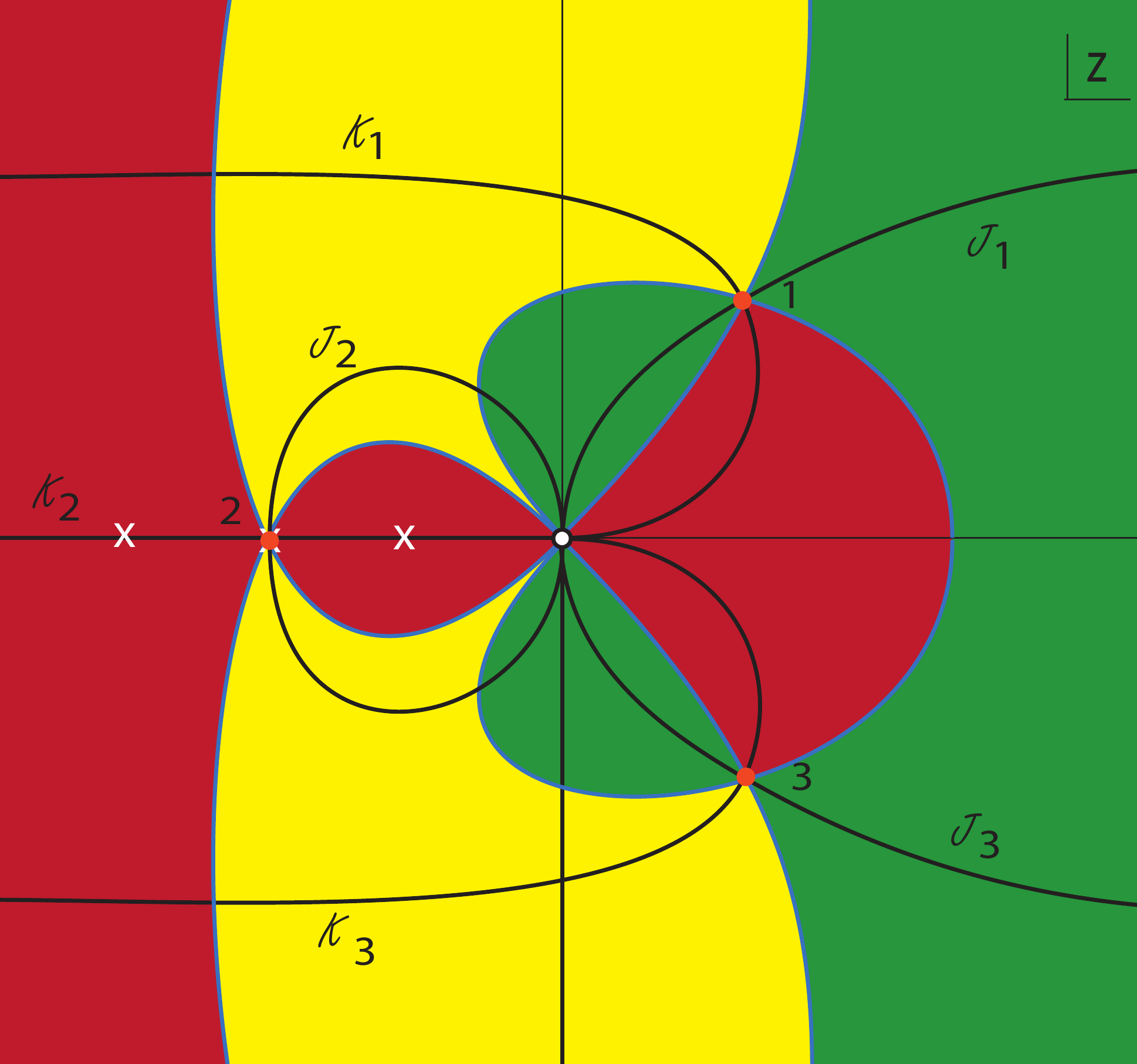}
\hspace{1cm}
\includegraphics[width=0.4\linewidth]{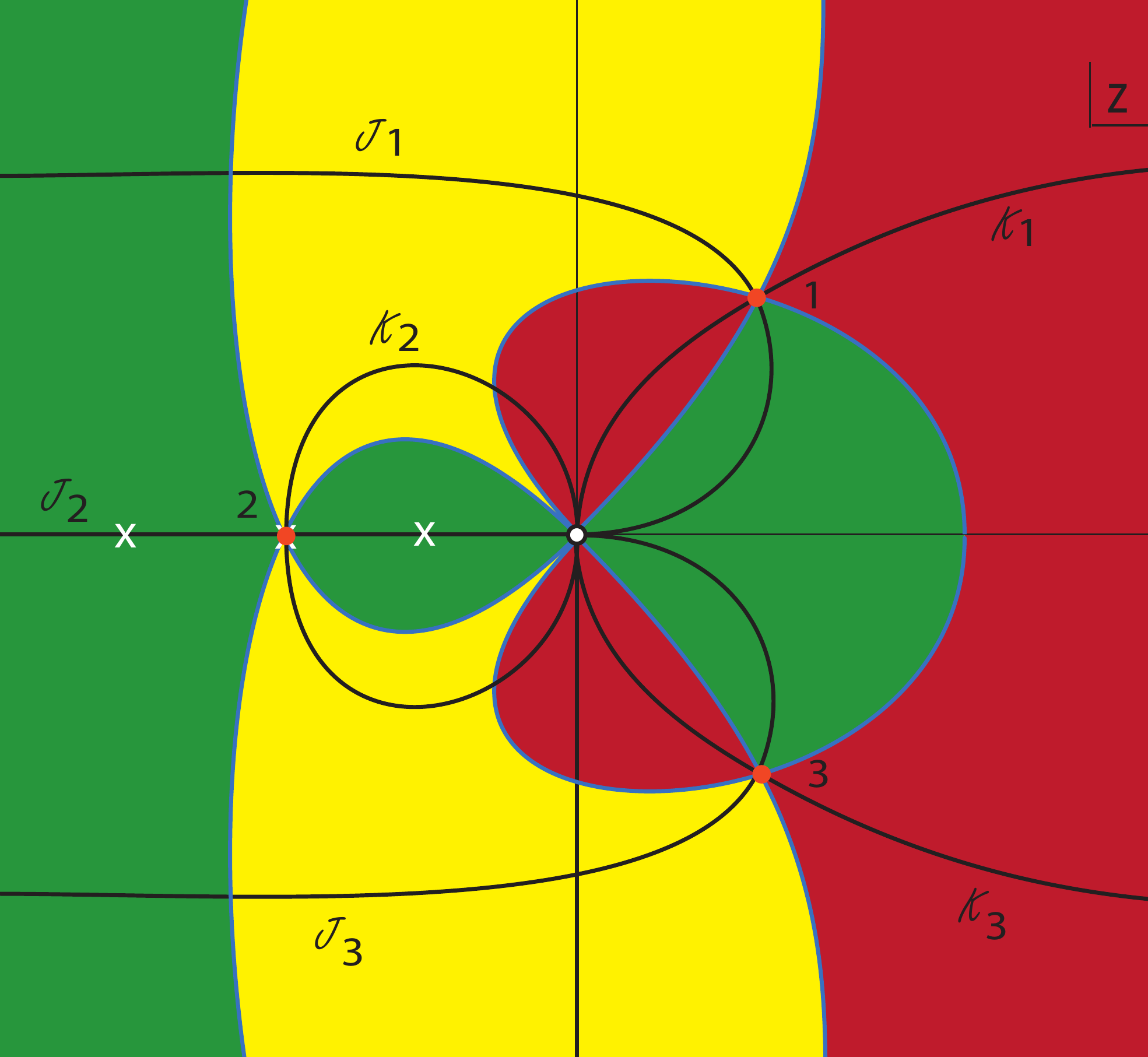}
\caption{The figure shows the qualitative structure of the Morse function in the complex $z$ plane, for boundary conditions  $1 < \frac{a_1^2 \Lambda}{3} \lessapprox 4.51.$ The left panel corresponds to $a(t) = - r \sin \left(\frac{N t}{r} \right)$, the right panel to the opposite choice of sign.  The flow lines are shown for the background. We will see that, when perturbations are added, poles  arise in the action (marked by white crosses here). A full description is provided in the main text.}
\label{fig:flows}
\end{center}
\end{figure}

\begin{figure}
\begin{center}
\includegraphics[width=0.4\linewidth]{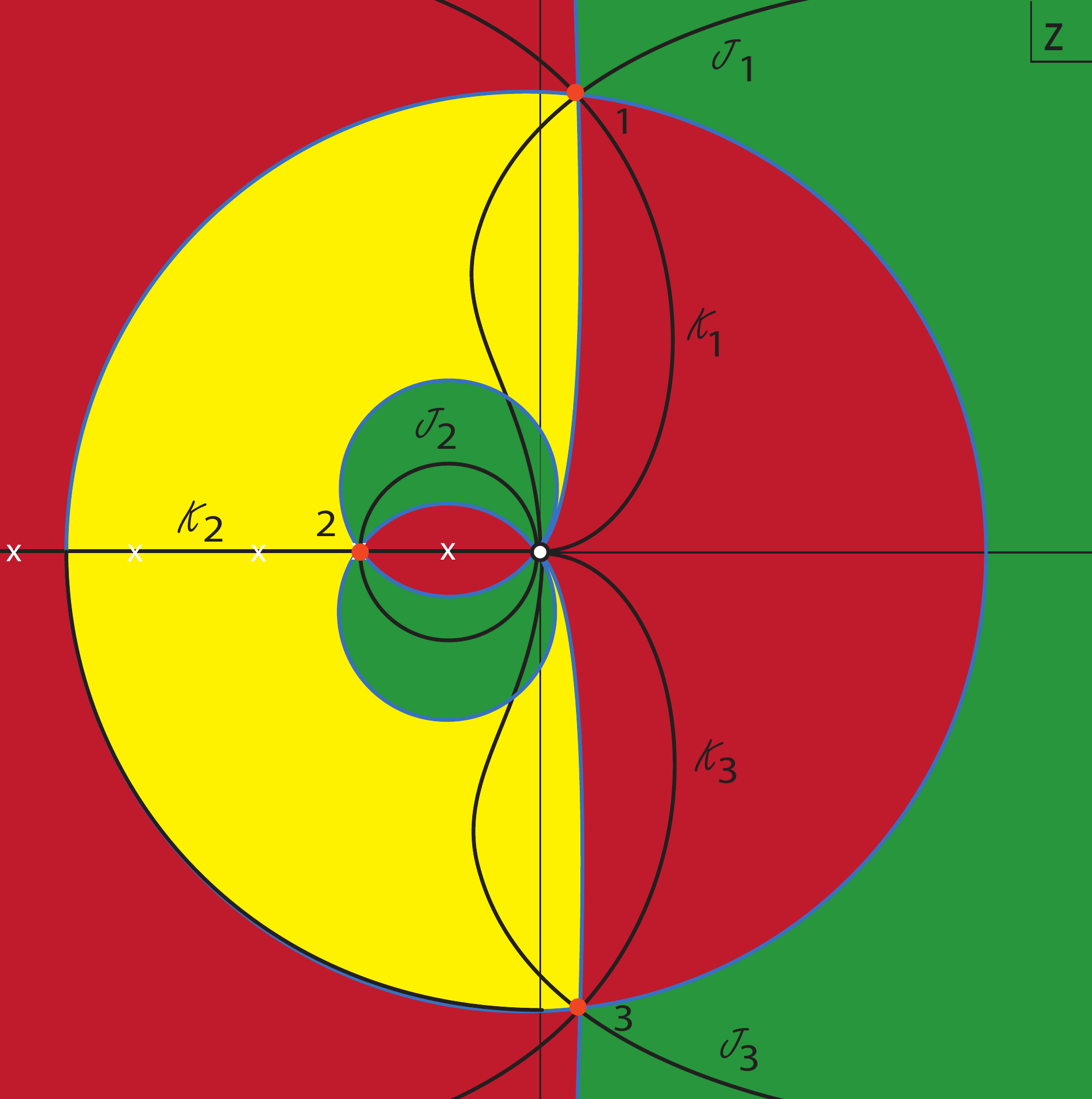}
\hspace{1cm}
\includegraphics[width=0.4\linewidth]{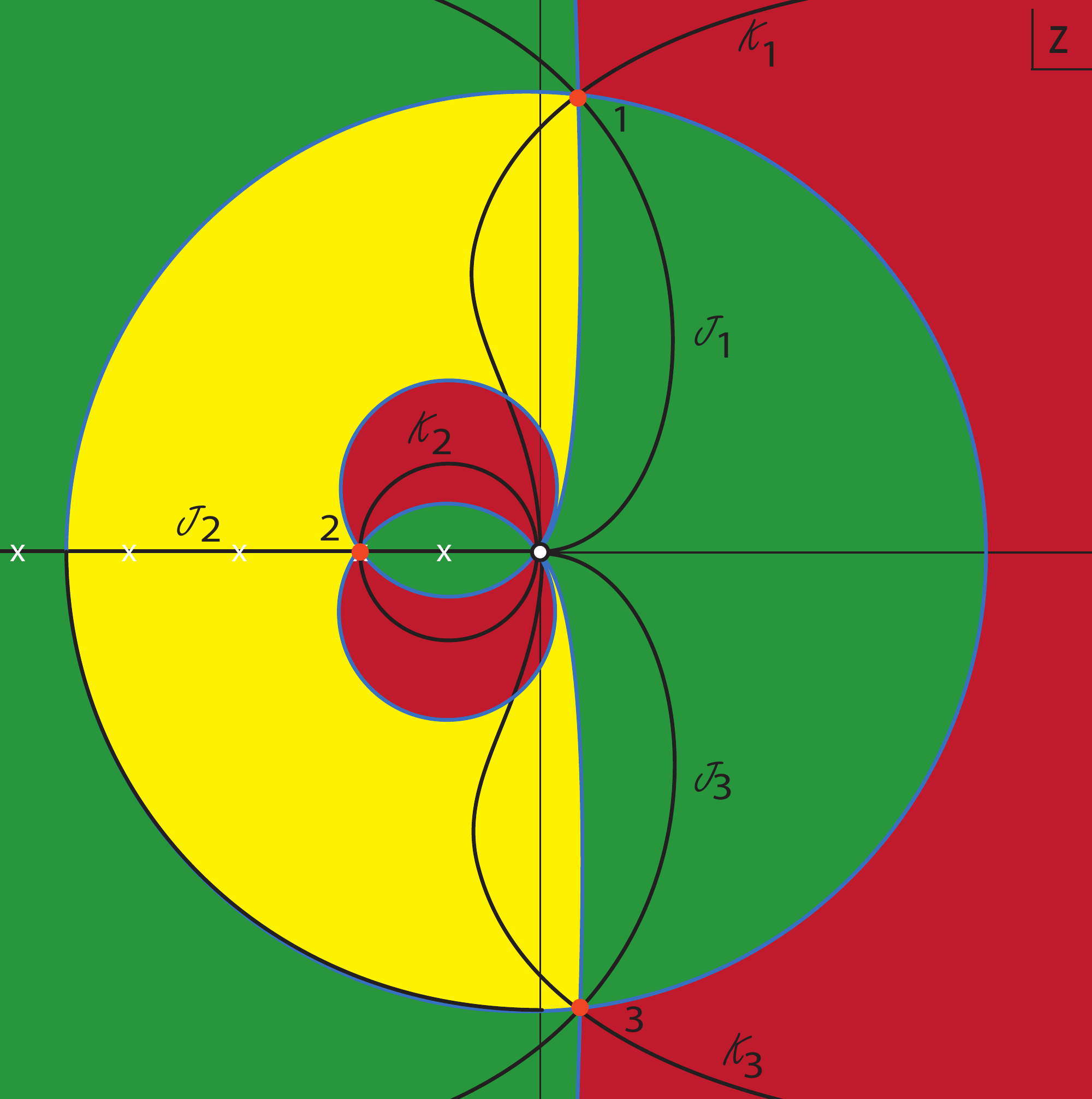}
\caption{The figure shows the qualitative structure of the Morse function in the complex $z$ plane, this time for boundary conditions  $\frac{a_1^2 \Lambda}{3} \gtrapprox 4.51.$  The left panel corresponds to $a(t) = - r \sin \left(\frac{N t}{r} \right)$, the right panel to the opposite choice of sign.  The flow lines are shown for the background and the locations of the poles, which arise when perturbations are added, are marked by white crosses here.}
\label{fig:flows2}
\end{center}
\end{figure}

The four saddle points at $z_{1,3}$ are those also seen in the minisuperspace calculation \cite{Feldbrugge:2017kzv}. These saddle points can be pictured as half of a 4-sphere glued onto half of the de Sitter hyperboloid, with a radius $r$ determined by the cosmological constant, $r^2 = \frac{3}{\Lambda},$ thus they are all four complex solutions to the Einstein equations. They differ in the way that the implied Wick rotation from the de Sitter geometry to the sphere is implemented, while two of the saddle points are the time reverses of the other two. In the present context these four saddle points arise for two possible sign choices for the complexified scale factor, whereas the minisuperspace calculation already includes a sum over both possible choices. By contrast, the saddle points at $z=-2$ are of a different character. As already discussed by Halliwell and Louko \cite{Halliwell:1989vu}, these are spurious solutions which do not satisfy the Einstein equations. Moreover, they do not lead to classical evolution, as they do not yield a phase in the exponent $e^{iS/\hbar}.$ 

Note that the position of the saddle points is the same for the two choices of sign for $a(t)$. However the value of the action is the opposite. As a consequence, the flow lines are the same, with exchanged roles of the steepest descent and ascent paths. This has important consequences for Picard-Lefschetz theory and the choice of integration contour. The locations of the saddle points in the complex $z$ plane, along with the paths of steepest ascent and descent emanating from them, are shown in Figs. \ref{fig:flows} (for small values of the final scale factor $1 < \frac{a_1^2 \Lambda}{3} \lessapprox 4.51$) and and \ref{fig:flows2} (for larger values of $a_1$).

We will now describe these figures -- for more details about the general procedure see \cite{Feldbrugge:2017kzv,Feldbrugge:2017mbc}. Figs. \ref{fig:flows} and  \ref{fig:flows2} show the qualitative behaviour of the Morse function, defined as the magnitude of the integrand. More specifically, one rewrites the integrand, now seen as a holomorphic function of the fields, as $e^{iS/\hbar} \equiv e^{h+iH},$ where $h, H$ are real functions. The Morse function $h$ then determines the amplitude of the integrand, while $H$ describes the phase. Critical points (in fact saddle points) of $h$ are also critical points of the total action, and the lines which have the same phase as that of a saddle point are the lines of steepest ascent/descent from that saddle point. Along these lines the Morse function changes most rapidly, and monotonically, away from the saddle points. The saddle points are marked by orange dots, the steepest ascent (${\cal K}$) and descent (${\cal J}$) lines are black, while the blue lines have the same value of the Morse function as the saddle points which they link up to. The green regions have a lower value of the Morse function than the adjacent saddle point, while the red regions have a higher Morse function. Yellow regions have a value of the Morse function in between the values at the two adjacent saddle points. The action has an essential singularity at $z=0.$ Approaching this within a green region thus implies a converging integral, while approaching it in a red region leads to divergence. Picard-Lefschetz theory aims to replace a highly oscillating integral by an equivalent absolutely convergent one along lines of steepest descent, if possible. The oscillating integral involves many cancellations due to the oscillations, while along a steepest descent line no such cancellations occur. Thus the Morse function along the steepest descent path must be lower than along the original integration contour. In other words, a steepest descent path (also called Lefschetz thimble) is relevant to the integral only of it can be reached by flowing the original integration contour down towards it. Equipped with these tools, we can discuss possible integration contours.

\begin{figure}
\begin{center}
\includegraphics[width=0.3\linewidth]{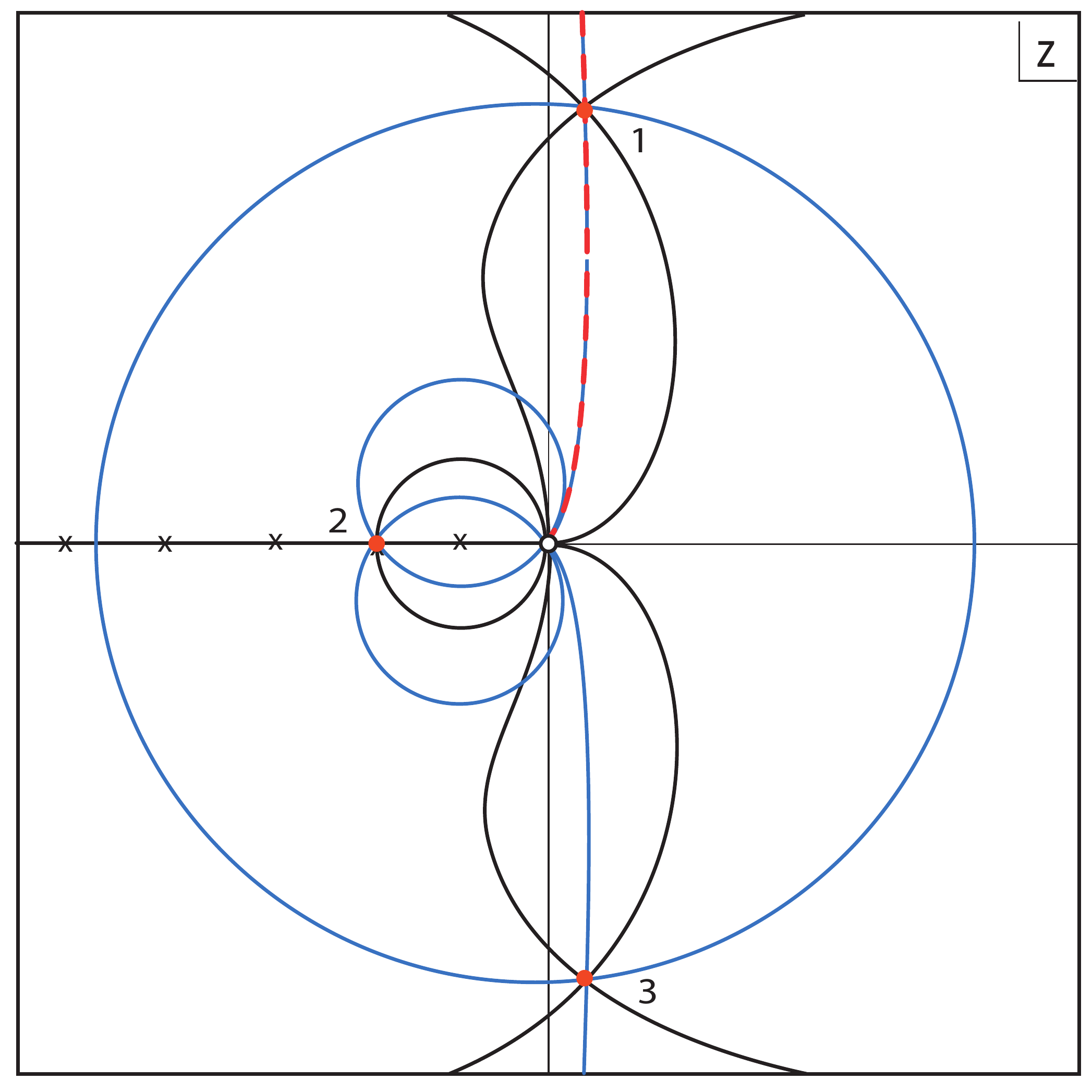}
\includegraphics[width=0.3\linewidth]{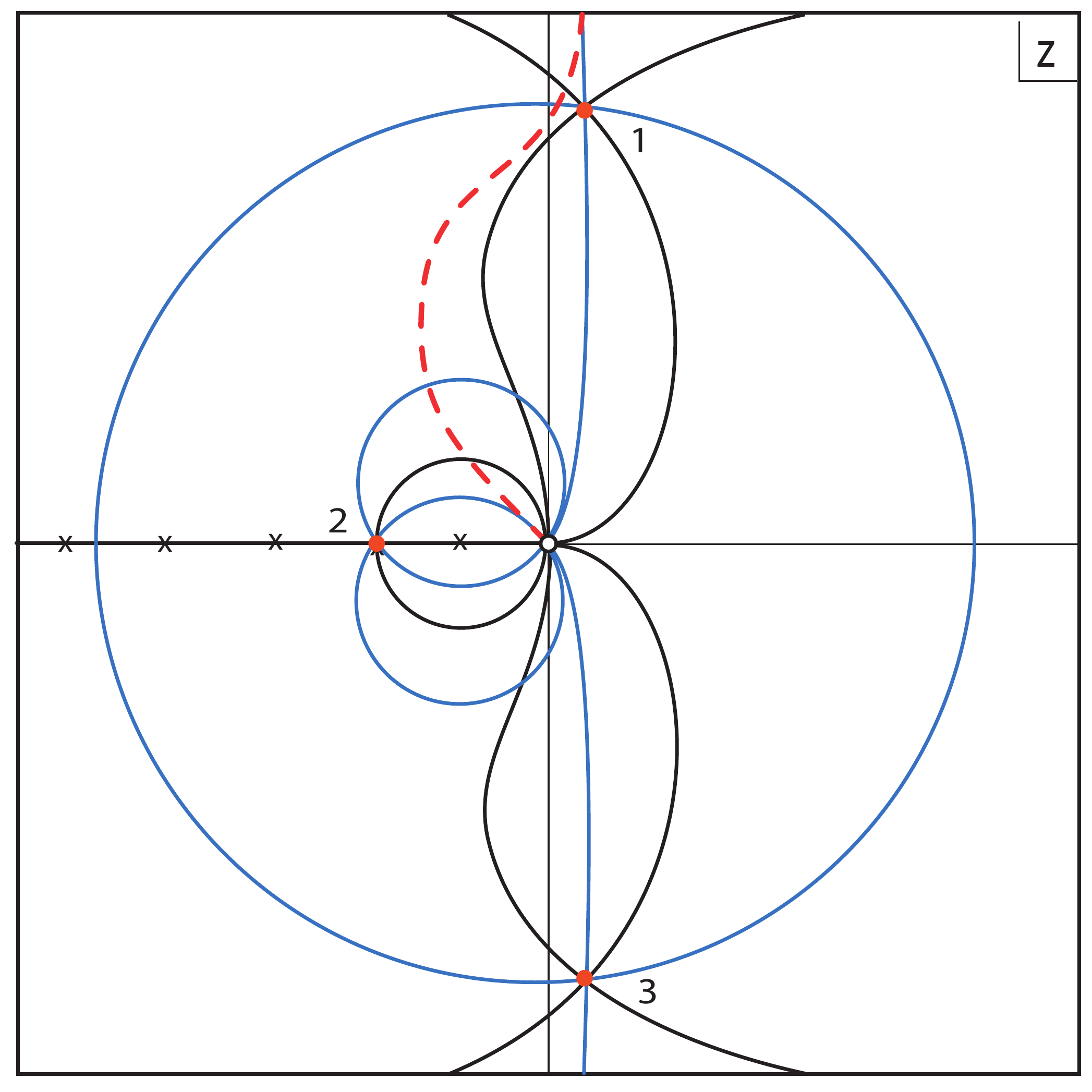}
\includegraphics[width=0.3\linewidth]{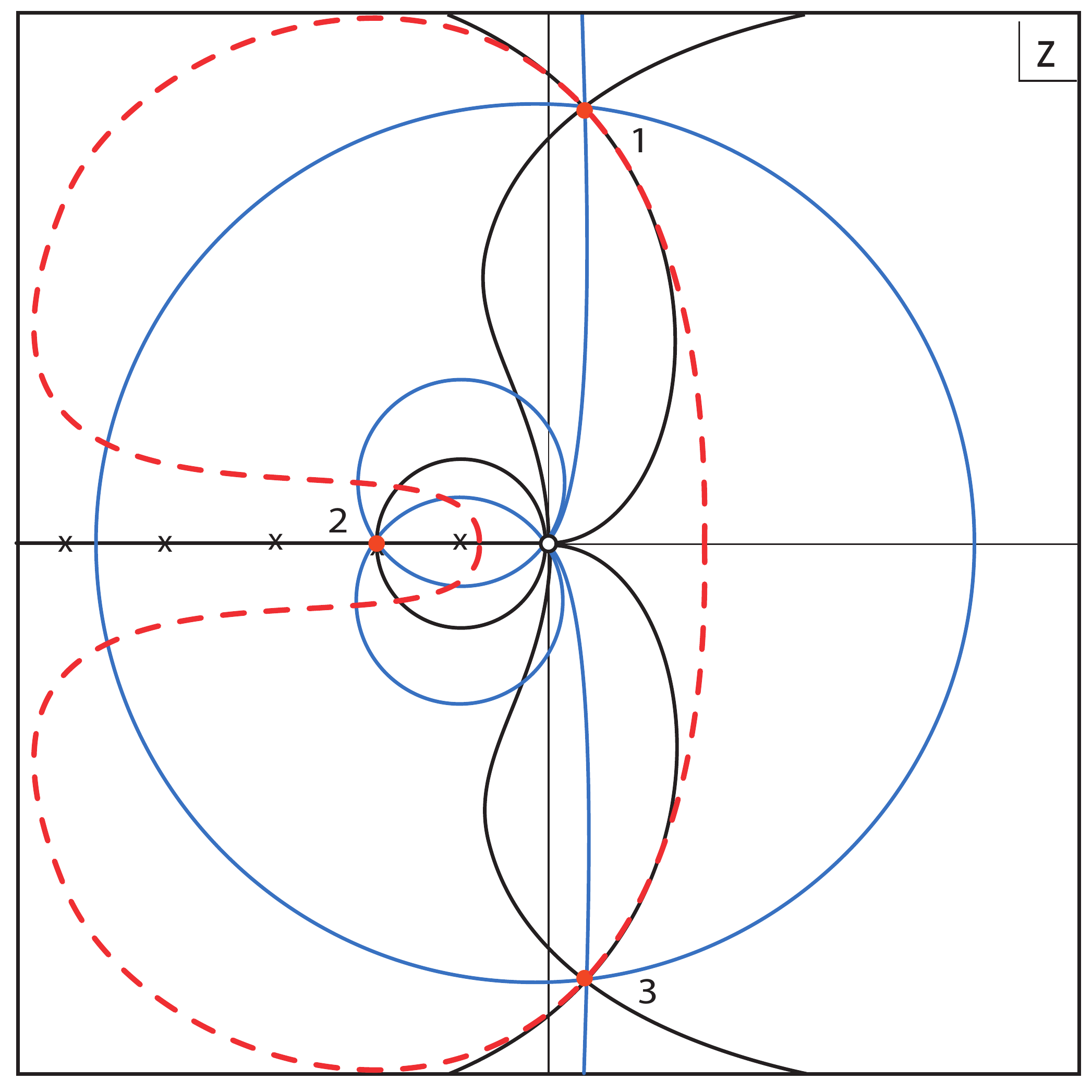}
\caption{Possible contours of integration are shown by the dashed orange lines. The figures here correspond to the physically most relevant boundary conditions where $\frac{a_1^2 \Lambda}{3} \gtrapprox 4.51.$ Analogous contours also exist for smaller values of $a_1.$}
\label{fig:contours}
\end{center}
\end{figure}

But before doing so, we should point out an important difference with minisuperspace calculations: the restricted version of the path integral considered here is in fact not straightforwardly related to the actual gravitational path integral. There, one defines the Lorentzian path integral so as to include a sum over all possible (real and positive) values of the lapse function $N_L$. This results in a Green function of the Wheeler-DeWitt (WdW) operator which possesses a well defined underlying causal structure \cite{Feldbrugge:2017kzv,Feldbrugge:2017mbc}. By contrast, the object constructed as a sum over regular metrics is not explicitly related to the WdW equation, as there is no explicit integration over the lapse function. As a consequence it is not obvious in this case what the appropriate integration contour should be.  In what follows we will define an appropriate integration contour by taking guidance from the gravitational path integral in order to identify desirable properties. First, the integration contour should run between singularities of the action. This implements the idea of summing over all possible configurations, since there is no reason to end the contour at any particular configuration along a path. Also, the convergence of the path integral with a finite boundary configuration highly depends on the measure. Since the measure is not uniquely defined here, we will rather consider endpoints where the integrand is exponentially suppressed, $iS \rightarrow - \infty$. Secondly, we will try to define an analogue of a Lorentzian contour. Indeed there is no contour in $z$ for which the 4-geometries have a purely Lorentzian signature. However, it is possible to require that the metrics become Lorentzian at late times, near the final boundary. This is equivalent to imposing that the velocity on the final boundary must be purely imaginary.  One can see this by rewriting the metric near the final boundary as
\begin{equation}
ds^2 = N_E^2 dt^2 + a(t)^2 d \Omega_3^2 = N_E^2 \left( dt^2 + \frac{a(t)^2}{N_E^2 \Delta t^2} \Delta t^2 d \Omega_3^2 \right)
\end{equation}
Indeed this line element has a Lorentzian signature near the final boundary if $\frac{\dot{a}_1}{N_E}$ is purely imaginary. According to the definition of $z,$ Eq. (\ref{z}), this requirement singles out the line $z = 1 + i s $ with $s$ real. Since opposite values of $s$ correspond to opposite final velocities, it is physically meaningful to restrict ourselves to the upper half complex $z$ plane, i.e. $s >0$. In fact, $s = 0 $ is a branch point for the map $r \rightarrow z$ (given by $z-1= \pm \sqrt{1-\frac{a_1^2}{r^2}}$) which must not be crossed for a proper definition of the $z$ variable. The line $1+i s$ can also be seen as the region where $r < a_1$. The appropriate integration contour should indeed approach this line for $ |z|\gg1$ (or equivalently when $r \ll a_1$). In this limit $N_E \rightarrow 0$ along the imaginary axis. Therefore the singularity of the action at infinity can be interpreted as the necessity of an ever bigger kinetic energy for a transition of the scale factor from zero to a finite value in ever smaller amounts of time. But where should the other end point of integration lie?  The only other singularity is at $z=0,$ hence it must lie there. By definition (see Eq.(\ref{z})), the limit $z \rightarrow 0 $ is equivalent to $ \cos \left( \frac{N_E}{r}\right) \rightarrow - 1$. Thus $ N_E \approx \pm \pi r$ there. Since  $z$ vanishes when $r \rightarrow \infty$, the lapse blows up in this limit, and thus the singularity at $z=0$ corresponds to the limit of infinite $r$ and $N_E.$ As an aside, note that $r$ also diverges near $z = 2$. However in this case there is no condition on the lapse, and the action remains perfectly finite there (in fact, for $z = 2$, $a(t) = \pm N_E t $ and $a_1 = \pm N_E$). 

In the present case, the end points alone are not enough to completely fix the contour of integration yet, as there are inequivalent directions of approach to the essential singularity at $z=0$. By analogy with the Lorentzian path integral, we will require our defining contours to lie in a region where the integral is conditionally convergent, i.e. a region that asymptotically borders the regions of manifest divergence and convergence. In Figs. \ref{fig:flows} and \ref{fig:flows2} the lines which asymptotically demarcate regions of convergence and divergence are the blue lines, and hence when approaching $z=0$ from above we have two choices: either approaching at an angle of $\pi/4$ or at an angle of $3\pi/4.$ The first possibility is shown in the left panel of Fig. \ref{fig:contours}, and consists of the dashed line passing through saddle point $1$. This line can be deformed into the thimble ${\cal J}_1$ for both choices of sign of the scale factor $a(t).$ If one were to sum over both choices of sign, then the resulting amplitude would, in the saddle point approximation, consist of a sum of the two saddle point contributions at $z_1,$ i.e. a sum of a suppressed saddle point with weighting $e^{-12\pi^2/(\hbar\Lambda)}$ and an enhanced saddle point with weighting $e^{+12\pi^2/(\hbar\Lambda)}.$ We will comment on this contour again at a later stage when discussing perturbations\footnote{We note that this contour would be related to the infinite complex contour proposed by Diaz Dorronsoro et al. in \cite{DiazDorronsoro:2017hti} -- in fact it would give ``half'' of that result. However, in minisuperspace a contour which would yield a similar result would not be Lorentzian at all, as it would have to run from $-i\infty$ to $+\infty$ in the complex plane of the lapse function, see e.g. Fig. $5$ in \cite{Feldbrugge:2017kzv}.}. For now, let us just re-iterate a comment already made in \cite{Feldbrugge:2017mbc}, which is that the enhanced saddle point does not obey the correspondence principle: in the limit that $\hbar \rightarrow 0,$ its weighting becomes larger and larger, so that this quantum effect becomes \emph{more} dominant in the classical limit, rather than less. This strongly suggests that the upper sign in Eq. \eqref{eq:summedmetrics} should in fact be discarded.

The second possibility of interest is shown in the middle panel of Fig. \ref{fig:contours}. Here the contour (orange dashed line) leaves the essential singularity along a blue line at an angle of $3\pi/4$ and asymptotically joins the ``Lorentzian'' $1+is$ line. This contour turns out to be the closest analogue of the Lorentzian contour in minisuperspace \cite{Feldbrugge:2017kzv}. For the choice $a(t)=-r\sin{N_E t/r},$ it is equivalent to the previous contour (as they are separated by a green region of convergence near $z=0,$ so that an arc linking the two contours at $z=0$ yields a vanishing contribution to the integral), see also the left panel of Fig. \ref{fig:flows2}. Again, it can be deformed into the thimble ${\cal J}_1$ and it will yield a propagator that can be approximated by the contribution of the saddle point at  $z_1,$ giving an amplitude $\propto e^{-12\pi^2/(\hbar\Lambda)}.$ This coincides with the result of the minisuperspace analysis. We will take this contour to be our preferred contour. For the opposite choice of sign for $a(t),$ this contour is inequivalent to the one in the left panel of Fig. \ref{fig:contours}, as can be seen very clearly in the right panel of Fig. \ref{fig:flows}. Now the two contours are separated by a red region of divergence near $z=0.$ The ``preferred'' contour now only crosses the ${\cal K}_2$ steepest ascent line, and consequently only the spurious (and in this case highly enhanced) saddle point $z_2$ contributes to the integral. (Moreover, as we will see below, the action for the perturbations develops a pole at $z=-2$ for perturbation modes with wavenumber $k=3$.) This unphysical result can be avoided by considering only the lower choice of sign in the sum over metrics \eqref{eq:summedmetrics}, in agreement with the comment made at the end of the last paragraph. With that restriction on the sum over metrics, both contours described above yield identical results.

\section{Perturbations at leading order} \label{sec:perts}

We are now in a position to add perturbations, i.e. we would like to evaluate the propagator
\begin{align}
G[a_1,\phi_1;0,0] = \int {\cal D}z\int {\cal D}\phi \, e^{iS^{tot}/\hbar}
\end{align}
where the total Lorentzian action, including a gravitational wave of wavenumber $k,$ amplitude $\phi$ and fixed polarisation, reads
\begin{equation}
S^{tot} = 2 \pi^2 \int_0^1 N_L dt \left[ - 3 \frac{a \dot{a}^2}{N_L^2} - \Lambda a^3 + 3 a  + \frac{a^3 \dot{\phi}^2}{2 N_L^2} - \frac{a}{2} (k^2 -1) \phi^2 \right]
\end{equation}
It would be straightforward to include sums over wavenumbers and polarisations, but in order to avoid clutter we omit this extension.

\begin{figure}
\includegraphics[width=0.45\linewidth]{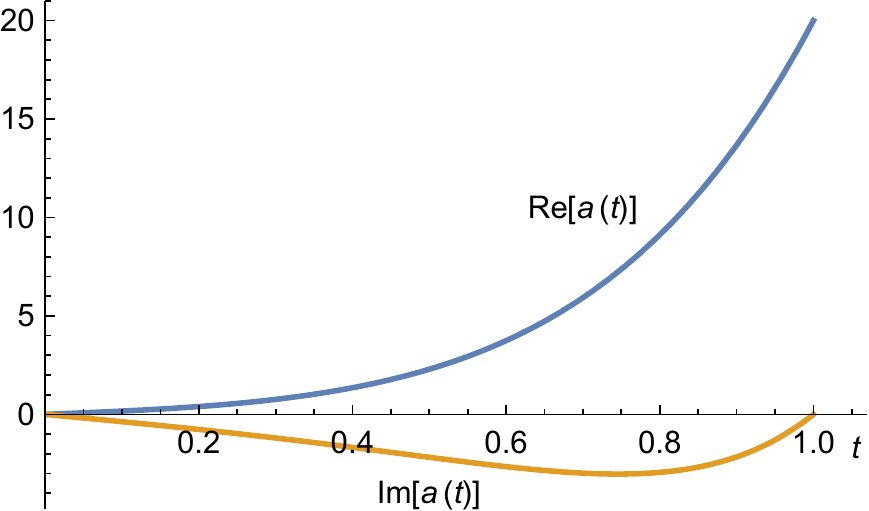}
\hspace{.5cm}
\includegraphics[width=0.45\linewidth]{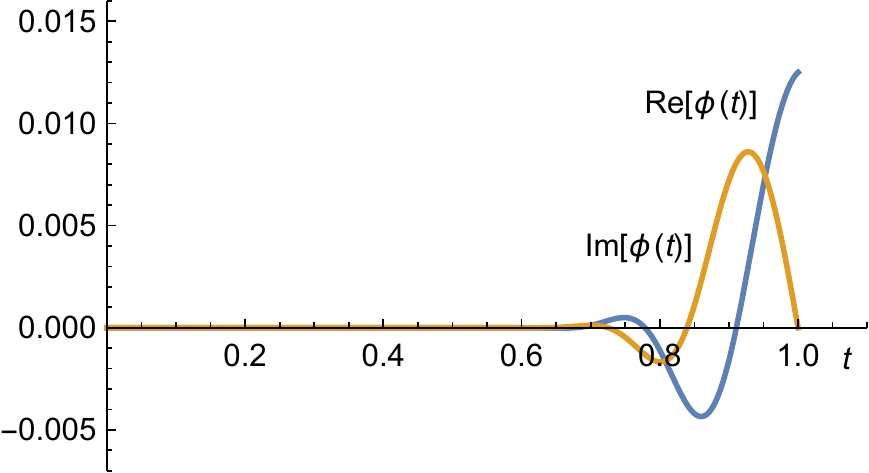}
\caption{The field evolution of the scale factor and of a gravitational wave perturbation at the saddle point $z_1 = 0.9997 + 19.97 i $ for the parameter values $\Lambda =3 $, $a_1 = 20 $, $\phi_1 = 1/80$, $k = 80$.}
\label{fig:fields}
\end{figure}

With a Euclidean lapse, the equation of motion for $\phi$ is given by
\begin{equation}
\frac{\ddot{\phi}}{N_E^2} + 3 \frac{\dot{a}}{a} \frac{\dot{\phi}}{N_E^2} - \frac{(k^2 - 1)}{a^2} \phi = 0   \label{phi}
\end{equation}
Note that this equation does not depend on the sign of $a(t)$. The general solution for both choices is given by \cite{Feldbrugge:2017fcc}
\begin{equation}
\begin{split}
F(t) = &+c_1 \frac{\left(1 - \cos \left( \frac{N_E t}{r}\right) \right)^{\frac{k-1}{2}} \left(\cos \left( \frac{N_E t}{r}\right)  + k\right)}{\left( 1 + \cos \left( \frac{N_E t}{r}\right)\right)^{\frac{k + 1}{2}}} +\\
  &+c_2  \frac{\left(1 + \cos \left( \frac{N_E t}{r}\right) \right)^{\frac{k-1}{2}} \left(\cos \left( \frac{N_E t}{r}\right)  - k\right)}{\left( 1 - \cos \left( \frac{N_E t}{r}\right)\right)^{\frac{k + 1}{2}}}\,,
\end{split}
\end{equation}
where $c_{1,2}$ are integration constants. The solution which is regular at $t =0 $ corresponds to setting $c_2 = 0,$ and thus the field perturbation is in fact zero at $a=0$. The boundary condition $\phi(t = 1) = \phi_1$ implies $c_1 = \frac{\phi_1}{F(1)}$ and $\phi(t) = c_1 F(t).$ An example of the field evolution is shown in Fig. \ref{fig:fields}.

The perturbative action reads
\begin{equation}
\begin{split}
S_{k}^{(2)} &=i \int_0^1 dt \left( \frac{a^3 \dot{\phi}^2}{2 N_E} + \frac{N_E}{2} a (k^2 - 1) \phi^2 \right)= \\
& = i\frac{d}{dt}\int_0^1 dt \, \frac{a^3 \phi \dot{\phi}}{N_E} - i\int_0^1 dt \frac{\phi(\ddot{\phi} + 3  H \dot{\phi})}{2 N_E} + i\int_0^1 dt \, N_E \frac{a}{2} (k^2 - 1) \phi^2
\end{split}
\end{equation}
Since the last two terms vanish along the solution to the e.o.m. for $\phi,$ the action takes the remarkably simple form
\begin{equation}
\begin{split}
S_{k}^{(2)} &= i\frac{a^3 \dot{\phi } \phi}{2 N_E} \mid_{t =  1} =i\frac{a_1^3 \dot{\phi_1 } \phi_1}{2 N_E}   = \pm i\frac{a_1^2 \phi_1^2}{2 } \frac{(k^2  -1) }{ \left[ \cos \left( \frac{N_E }{r}\right)+ k \right] }\\ & = \pm i\frac{a_1^2}{2} \frac{(k^2 - 1)}{(z + k - 1)} \phi_1^2 \label{eq:actionpert}
\end{split}
\end{equation}
The perturbative action evaluated at the saddle points reads respectively 
\begin{align}
S_{k}^{(2)}(z_2) &= \pm i\frac{(k^2 -1 ) \phi_1^2}{2 (k -3)} \\
S_{k}^{(2)} (z_1) & = \pm i\frac{a_1^2 (k^2 - 1)}{2 (k + i \sqrt{\frac{a_1^2 \Lambda}{3} - 1})} \phi_1^2 \, \approx\,  \pm \frac{\sqrt{3} (k^2 - 1) a_1 \phi_1^2}{2 \sqrt{\Lambda}} \pm i\frac{3k(k^2-1)\phi_1^2}{2\Lambda} \label{actionz1}\\
S_{k}^{(2)} (z_3) & = \pm i\frac{a_1^2 (k^2 - 1)}{2 (k - i \sqrt{\frac{a_1^2 \Lambda}{3} - 1})} \phi_1^2 \, \approx \, \mp \frac{\sqrt{3} (k^2 - 1) a_1 \phi_1^2}{2 \sqrt{\Lambda}} \pm i\frac{3k(k^2-1)\phi_1^2}{2\Lambda}\,,
\end{align}
where the approximate expressions correspond to the limit of a large final scale factor value $a_1.$ 

Two properties stand out immediately: first, the implied weighting of the perturbations is Gaussian for the upper sign (i.e. also for the upper sign in Eq. \eqref{eq:summedmetrics}), and inverse Gaussian ($\sim e^{+k^3\phi_1^2/{\hbar \Lambda}}$) for the lower sign. Since the Lorentzian saddle point corresponds to the lower sign, we find that the ``preferred''  contour, i.e. the one that coincides most closely with the Lorentzian minisuperspace contour, yields a propagator of the form
\begin{align}
G^{Lorentz}[a_1,\phi_1;0,0] \approx e^{- \frac{ 12 \pi^2 }{\hbar\Lambda} \left[1 + i \left(\frac{a_1^2 \Lambda}{3} -1 \right)^{3/2}\right]  + \frac{3k(k^2-1)\phi_1^2}{2\hbar\Lambda} - \frac{i \sqrt{3} (k^2 - 1) a_1 \phi_1^2}{2 \hbar\sqrt{\Lambda}}} 
\end{align}
The present calculation confirms that the no-boundary transition amplitude from nothing to a large final universe is unstable, even when only regular geometries are summed over. This is our main result. It demonstrates that the instability of the no-boundary proposal is not due to off-shell singularities, but is an intrinsic feature of the Lorentzian no-boundary/tunneling saddle points. 

\begin{figure}
\begin{center}
\includegraphics[width=0.45\linewidth]{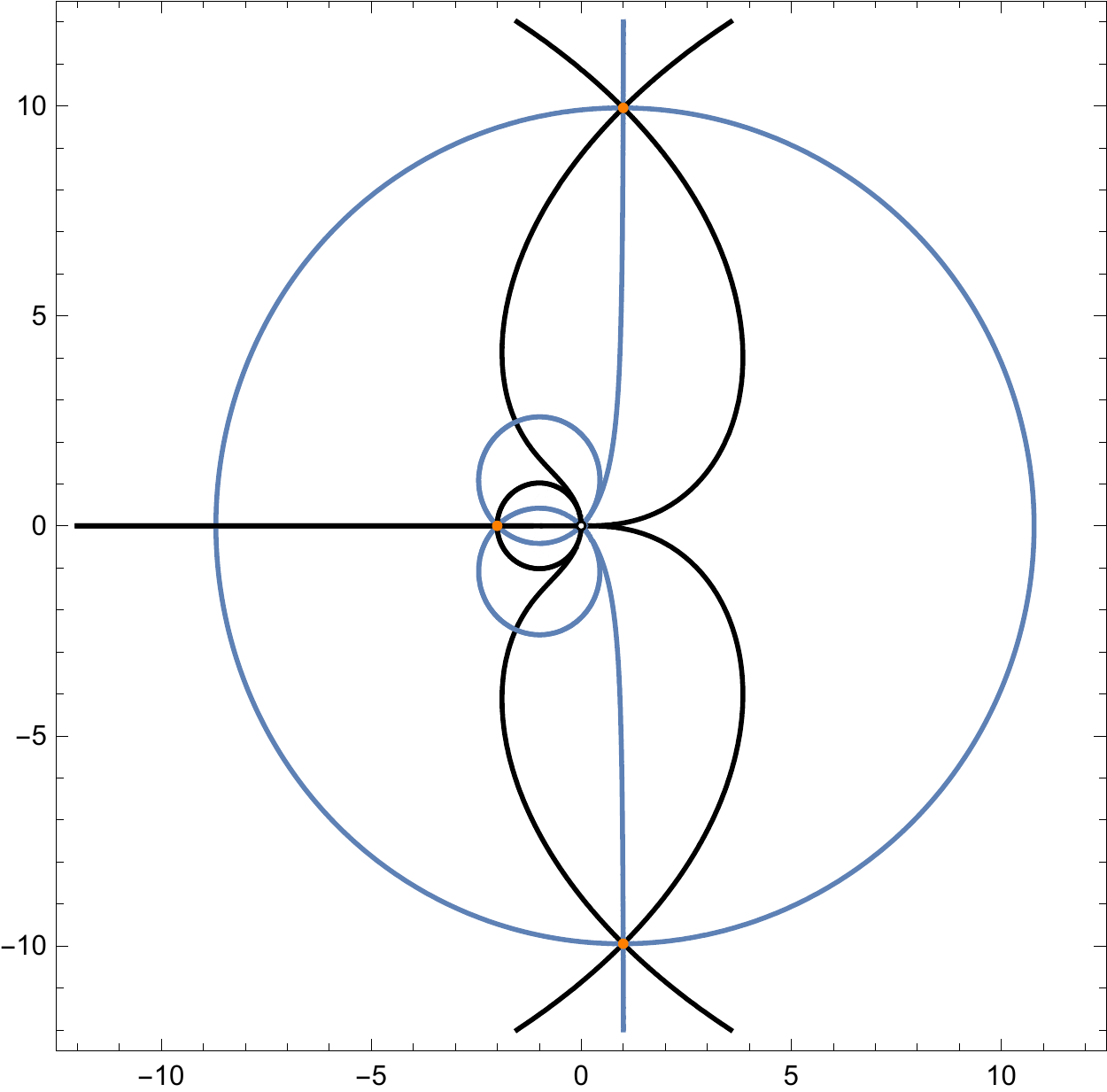}
\hspace{.5cm}
\includegraphics[width=0.45\linewidth]{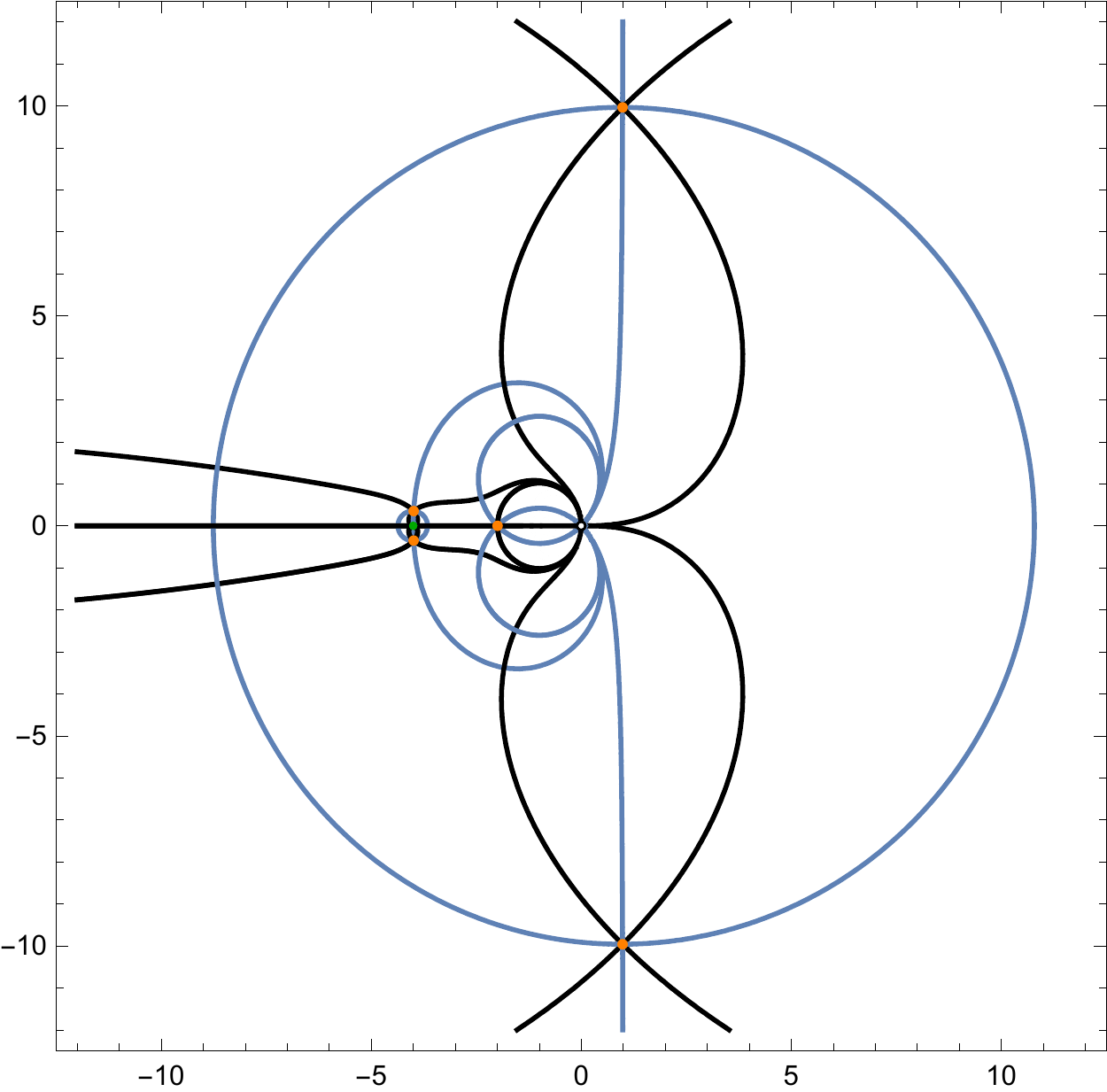}
\caption{Left panel:  flow lines for large values of the final scale factor, cf. Fig. \ref{fig:flows2}. In the figure the parameters are $\Lambda=3, a_1 = 10.$ Saddle points are marked in orange, black lines are steepest ascent/descent lines (the directions of descent, say, can be inferred straightforwardly by comparison with the two panels in Fig. \ref{fig:flows2}), and blue lines have the same value of the Morse function as the saddle points that they cross. Right panel: adding a gravitational wave perturbation of wavenumber $k$ results in two additional saddle points near the new pole of the action at $z=-k+1$. Here the parameters that were used are $\Lambda=3, a_1=10, k=5, \phi_1=1/5,$ with the pole (in green) located at $z=-4.$ The figure shows that the new flow lines, associated with the two additional saddle points, are irrelevant for our preferred contour of integration.}
\label{fig:contourslargea1}
\end{center}
\end{figure}

The second feature is that the perturbative action contains a wavenumber-dependent pole at $z=-k+1,$ as is evident form Eq. \eqref{eq:actionpert}.  One consequence is that two new saddle points appear near this pole with $Re(z_{saddle}) < 0$. The flow lines associated to those saddle points end at the new pole $ z = 1 - k$ and otherwise remain close to the negative $z$ axis -- see Fig. \ref{fig:contourslargea1} for an illustration. As $k$ increases, the pole moves away from the origin along the negative real line. The contours discussed so far receive no contribution from these new saddles, while the positions of their relevant saddle points are merely shifted by negligible amounts. However, any contour that crosses the negative real $z$ line is liable to non-trivial corrections. In particular, a circular contour such as that proposed in \cite{DiazDorronsoro:2018wro} becomes essentially untenable: for it to be well defined, it must encircle the origin at a radius smaller than the closest pole. Given that small $|z|$ corresponds to large $r,$ this would imply that a circular contour can only sum over very large $r$ geometries (which do not contain a Lorentzian region near the final boundary), i.e. certainly not over a representative set of regular geometries. In addition such a contour could not be deformed into the two thimbles ${\cal J}_{(1,2)},$ as envisioned in \cite{DiazDorronsoro:2018wro}, but would receive additional (wavenumber-dependent) contributions from the perturbative poles, as illustrated in the right panel of Fig. \ref{fig:contours}. This type of contour is thus of no physical interest, as argued before in \cite{Feldbrugge:2018gin}.

A further consequence concerns the contour presented in the left panel of Fig. \ref{fig:contours}. With this choice of contour, both signs of the scale factor are allowed, and the propagator becomes a sum over both contributions. This propagator then consists of a stable and an unstable saddle point, and for large $a_1$ it has a weighting of the form
\begin{align}
|G[a_1,\phi_1;0,0]| \approx e^{-\frac{ 12 \pi^2 }{\hbar\Lambda} + \frac{3k(k^2-1)\phi_1^2}{2\hbar\Lambda}} + e^{+ \frac{ 12 \pi^2 }{\hbar\Lambda} - \frac{3k(k^2-1)\phi_1^2}{2\hbar\Lambda}} 
\end{align}
For large enough wavenumbers $k$ and final amplitudes $\phi_1$, the unstable part dominates and the overall probability distribution becomes unstable. Subject to verifying that this result resides within the limits of linear perturbation theory, this confirms the results of \cite{Feldbrugge:2017mbc} obtained in minisuperspace.

This last observation brings us to the subject of backreaction, as it is important to know the range of validity of linear perturbation theory. In order to discuss backreaction, it is useful to re-iterate the calculational strategy employed in the present work: our aim is to sum over regular geometries. These are simply off-shell geometries, chosen with the unique criterion that they should be regular. For the background, we look at a particularly simple sub-class, namely 4-spheres with varying radii (or even just a subset of those, given the choice of sign of the scale factor). A priori these geometries do not satisfy any equations of motion, they are just a particular subset of geometries that are summed over in the path integral. Then we add perturbations, subject to two criteria: they should not destroy the regularity, and they should satisfy the linear equation of motion around the off-shell background geometries. Hence one should think of them as saddle points of the $\phi$ integral. Only after both integrals over $\phi$ and $z$ have been performed do we expect the final saddle point to be a solution to the full Einstein equations. Thus it only makes sense to check backreaction at the final saddle point.  This is different for the minisuperspace case studied in \cite{Feldbrugge:2017mbc}, where the lapse integral was over geometries that were already saddle points of the integral over the scale factor $q$. Hence each such configuration was already a solution to the $q$ equation of motion, and it made sense to check whether there was a large backreaction or not on those solutions, before the lapse integral was performed. But in the present context, we are only interested in whether the final overall saddle point is trustworthy.

The action for the background and perturbations leads to a system of two coupled differential equations, where the equation of motion for the scale factor $a(t)$ is modified, compared to the background, by $\phi$-dependent terms 
\begin{equation}
- 2 \frac{\ddot{a}}{a N_E^2} - \frac{\dot{a}}{a^2 N_E^2 } + \frac{1}{a^2} = \Lambda + \frac{1}{2 }\frac{\dot{\phi}^2}{N_E^2} + \frac{(k^2 - 1)}{6 a^2} \phi^2 \label{back1}
\end{equation}
Absence of backreaction corresponds to neglecting the kinetic and the gradient terms for $\phi$ in this equation. Thus a conservative view is to demand that these additional terms remain small at every value of $t$, i.e. that
\begin{equation}
|\frac{1}{2 }\frac{\dot{\phi}^2}{N_E^2}| , \, |\frac{(k^2 - 1)}{6 a^2} \phi^2| \, \ll \Lambda \qquad  \forall t \in [ 0 , 1] \label{back2}
\end{equation}

\begin{figure}
\begin{center}
\includegraphics[width=0.45\linewidth]{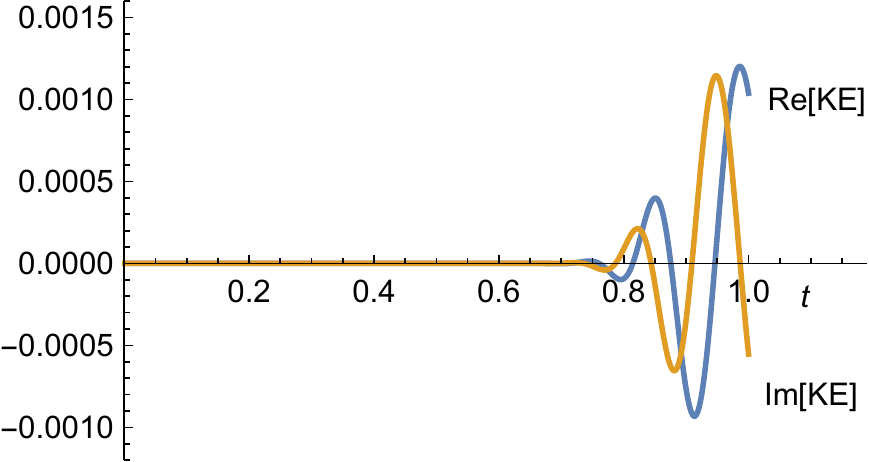}
\hspace{.5cm}
\includegraphics[width=0.45\linewidth]{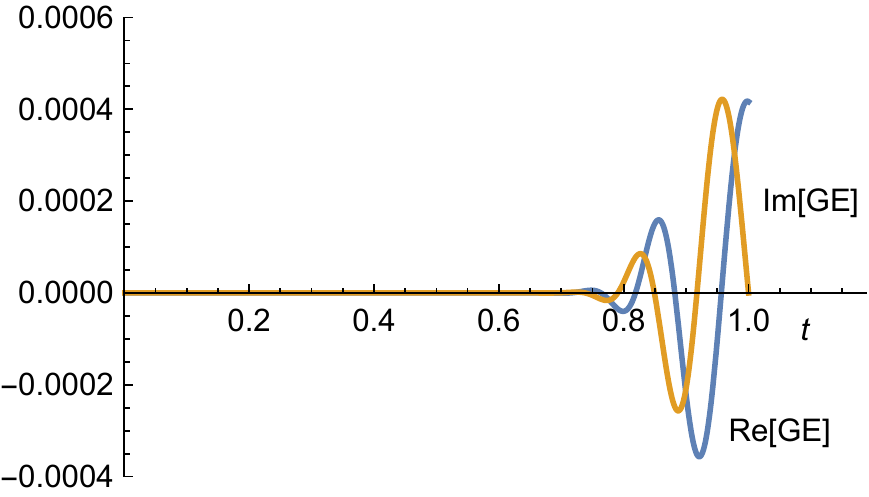}
\caption{The figure shows the behaviour of the backreaction terms involving the kinetic energy (KE)  $ \frac{1}{2 }\frac{\dot{\phi}^2}{N_E^2} $  (left panel) and the gradient energy (GE) $ \frac{(k^2 - 1)}{6 a^2} \phi^2$  (right panel) at the relevant saddle point  for $ \Lambda = 3, a_1 = 20, k=80, \phi_1 = 1/80$ (the field evolutions were shown in Fig. \ref{fig:fields}). These terms are everywhere small compared to $\Lambda$. }
\label{fig:backreaction}
\end{center}
\end{figure}

For large $k$ sub-Hubble modes, i.e. for modes that obey $k > a_1 \sqrt{\Lambda/3},$ the perturbation grows fastest right at the end, near $t=1.$ Thus the backreaction is also largest there, see for instance the numerical example in Fig. \ref{fig:backreaction}. Still, the backreaction remains negligibly small, staying well below a tenth of a percent at all times for this example where the parameters that are used satisfy $k \phi_1 = 1,$ so that the contribution of the perturbations to the total weighting is in fact large (it is larger than that of the background). We may in fact find analytic expressions for the backreaction at $t=1,$ using the results of the calculation of the perturbative action in Eq. \eqref{eq:actionpert},
\begin{align}
\frac{\dot\phi^2}{2N_E^2}(t=1) = \frac{(k^2-1)^2}{2 (z_1+k-1)^2} \left( \frac{\phi_1}{a_1}\right)^2\,, \qquad \frac{(k^2 - 1)}{6 a^2} \phi^2 (t=1) = \frac{(k^2 - 1)}{6} \left( \frac{\phi_1}{a_1}\right)^2\,.
\end{align}
The backreaction at $t=1$ scales as $\left( \frac{k \phi_1}{a_1}\right)^2,$ and it will be small compared to $\Lambda$ as long as 
\begin{align}
\phi_1 \ll \frac{a_1}{k \sqrt{\Lambda}}\,,\, \qquad \left( k > a_1 \sqrt{\frac{\Lambda}{3}}\right) \,.
\end{align}
Note that this bound does not preclude a large contribution from the perturbative action \eqref{actionz1} to the total weighting.

\begin{figure}
\begin{center}
\includegraphics[width=0.32\linewidth]{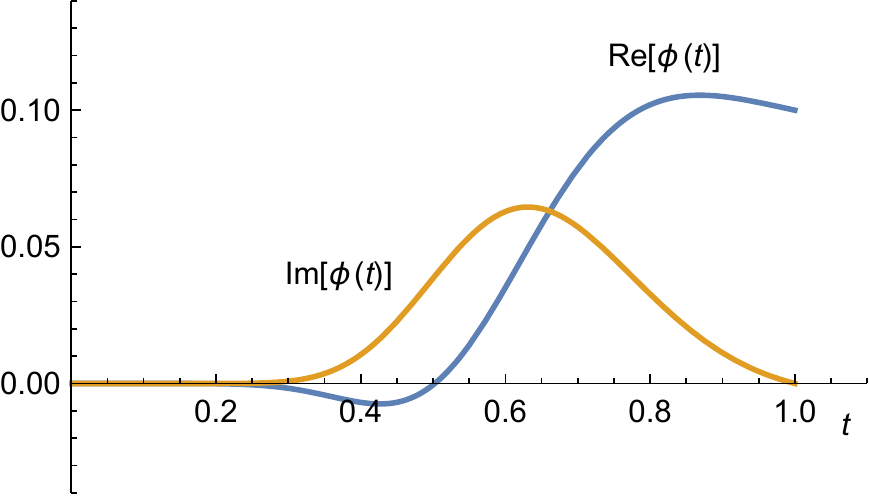}
\includegraphics[width=0.32\linewidth]{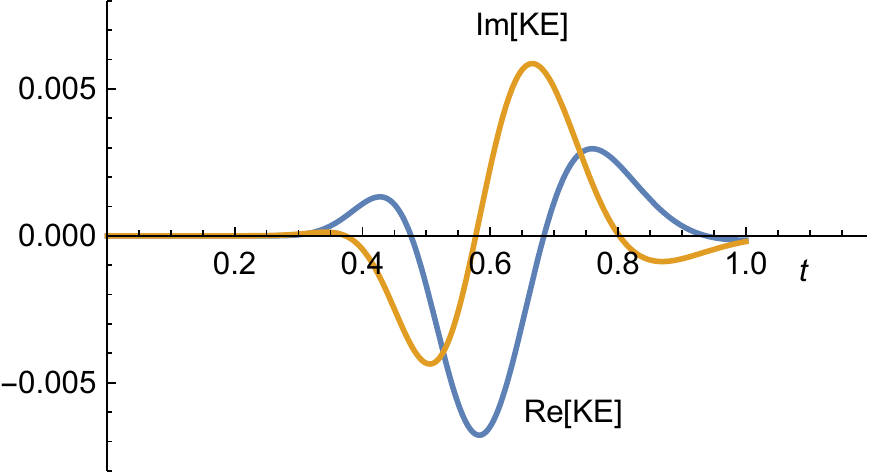}
\includegraphics[width=0.32\linewidth]{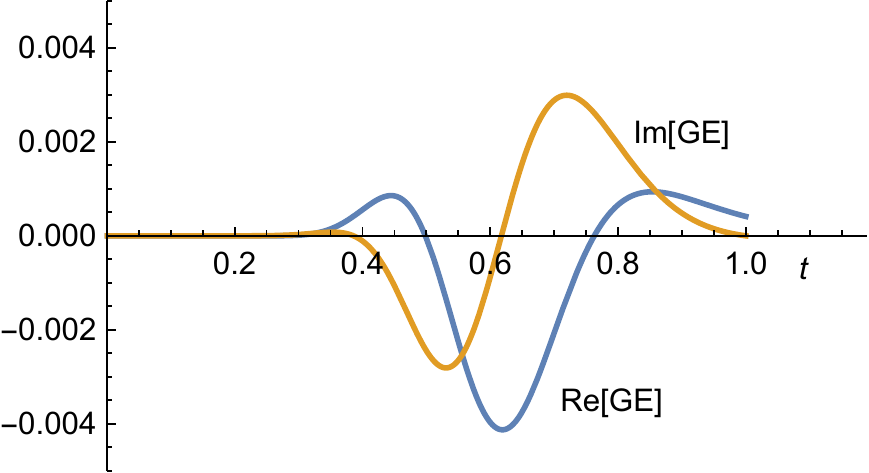}
\caption{The field evolution and backreaction terms for a perturbation with $ \Lambda = 3, a_1 = 20, k=10, \phi_1 = 1/10.$}
\label{fig:backreaction10}
\end{center}
\end{figure}

For super-Hubble modes, with $k \lesssim a_1 \sqrt{\Lambda/3},$ the mode functions grow earlier and then freeze out. Thus their main contribution to backreaction occurs significantly before the final hypersurface is reached. A numerical example is shown in Fig. \ref{fig:backreaction10},  where the same background was used as above, but now with wavenumber $k=10$ and final amplitude $k=1/10.$ Again, one can see that the backreaction terms stay below two tenths of a percent, as compared to the cosmological term $\Lambda=3.$ For smaller $k$ we find that the maximally allowed amplitude to stay within the regime of validity of linear perturbation theory does not decrease, such that overall linear perturbation theory is a very good approximation for a large parameter range.

\section{Discussion}

The quest to find a theory of initial conditions for the universe is intimately related to the quest of learning how to apply quantum theory to the universe. An appealing framework in this regard is the path integral approach to gravity. While this formulation is in all likelihood intrinsically limited to the semi-classical domain, it is highly useful since it provides a direct link to non-gravitational formulations of quantum theory, and since it allows one to use geometrical intuitions and methods.  Out of these geometrical considerations arose the no-boundary proposal of Hartle and Hawking, which may be seen as a proposal for replacing the big bang by sums over compact, regular geometries. By construction, the curvature singularity at the big bang is then avoided, and one may hope to find finite, well-defined results, ideally in agreement with observations.

A difficulty has been to define the gravitational path integral precisely, in particular to properly define the class of geometries that should be summed over. Mostly, interest has focussed on minisuperspace models, which still contain many singularities off-shell, but where the saddle points correspond to the smooth geometries that Hartle and Hawking had in mind. Unfortunately, a mathematically precise implementation of the no-boundary idea in terms of a Lorentzian path integral resulted in the conclusion that perturbations around these geometries are unstable, and that the proposal is untenable as a theory of the initial conditions of the universe, at least in the context of a universe dominated by a positive cosmological vacuum energy. A question which arose from this work was whether the instability was in fact caused by these off-shell singularities. For this reason, we investigated the restriction to summing over only manifestly regular geometries, in the simplest possible case of summing only over complexified 4-spheres of various radii. This approach was pioneered by Halliwell and Louko, and we extended their calculations by finding a complementary set of 4-spheres to be summed over, and by adding perturbations. Our calculations confirm that the no-boundary condition leads to an instability, well within the limits of applicability of linear perturbation theory.  

Thus we can conclude that the reason for the failure of the no-boundary proposal was not the breakdown of perturbation theory, nor the inclusion of off-shell geometries that included singular configurations and led to a non-analytic structure of the perturbative action (which incidentally, might be avoided by performing the minisuperspace path integral using Picard-Lefschetz theory extended to the infinite-dimensional case). However, in our view one should keep an open mind regarding other applications. It may still prove to be a fruitful idea to sum over regular (but complexified) metrics only, with different dynamics, different field content and for other types of boundary condition. After all, it is in no way clear yet that the universe had to be dominated by a large positive vacuum energy at its birth. Furthermore, there exist other singularities, in particular those in the interiors of black holes, that it might be interesting to investigate along these lines.

\acknowledgments

We would like to thank Sebastian Bramberger, Job Feldbrugge, Jonathan Halliwell and Neil Turok for many illuminating discussions.

\bibliographystyle{utphys}
\bibliography{RegularMetrics}

\providecommand{\href}[2]{#2}\begingroup\raggedright\begin{thebibliography}{10}

\bibitem{Bars:2011aa}
I.~Bars, S.-H. Chen, P.~J. Steinhardt, and N.~Turok, ``{Antigravity and the Big
  Crunch/Big Bang Transition},''
  \href{http://dx.doi.org/10.1016/j.physletb.2012.07.071}{{\em Phys. Lett.}
  {\bf B715} (2012)  278--281},
\href{http://arxiv.org/abs/1112.2470}{{\tt arXiv:1112.2470 [hep-th]}}.
%%CITATION = ARXIV:1112.2470;%%.

\bibitem{Boyle:2018tzc}
L.~Boyle, K.~Finn, and N.~Turok, ``{$CPT$ symmetric universe},''
\href{http://arxiv.org/abs/1803.08928}{{\tt arXiv:1803.08928 [hep-ph]}}.
%%CITATION = ARXIV:1803.08928;%%.

\bibitem{Garriga:1997ef}
J.~Garriga and A.~Vilenkin, ``{Recycling universe},''
  \href{http://dx.doi.org/10.1103/PhysRevD.57.2230}{{\em Phys. Rev.} {\bf D57}
  (1998)  2230--2244},
\href{http://arxiv.org/abs/astro-ph/9707292}{{\tt arXiv:astro-ph/9707292
  [astro-ph]}}.
%%CITATION = ASTRO-PH/9707292;%%.

\bibitem{Steinhardt:2001st}
P.~J. Steinhardt and N.~Turok, ``{Cosmic evolution in a cyclic universe},''
  \href{http://dx.doi.org/10.1103/PhysRevD.65.126003}{{\em Phys. Rev.} {\bf
  D65} (2002)  126003},
\href{http://arxiv.org/abs/hep-th/0111098}{{\tt arXiv:hep-th/0111098
  [hep-th]}}.
%%CITATION = HEP-TH/0111098;%%.

\bibitem{Lehners:2009eg}
J.-L. Lehners, P.~J. Steinhardt, and N.~Turok, ``{The Return of the Phoenix
  Universe},'' \href{http://dx.doi.org/10.1142/S0218271809015977}{{\em Int. J.
  Mod. Phys.} {\bf D18} (2009)  2231--2235},
\href{http://arxiv.org/abs/0910.0834}{{\tt arXiv:0910.0834 [hep-th]}}.
%%CITATION = ARXIV:0910.0834;%%.

\bibitem{Banks:1996vh}
T.~Banks, W.~Fischler, S.~H. Shenker, and L.~Susskind, ``{M theory as a matrix
  model: A Conjecture},''
  \href{http://dx.doi.org/10.1103/PhysRevD.55.5112}{{\em Phys. Rev.} {\bf D55}
  (1997)  5112--5128},
\href{http://arxiv.org/abs/hep-th/9610043}{{\tt arXiv:hep-th/9610043
  [hep-th]}}.
%%CITATION = HEP-TH/9610043;%%.

\bibitem{Gielen:2013naa}
S.~Gielen, D.~Oriti, and L.~Sindoni, ``{Homogeneous cosmologies as group field
  theory condensates},'' \href{http://dx.doi.org/10.1007/JHEP06(2014)013}{{\em
  JHEP} {\bf 06} (2014)  013},
\href{http://arxiv.org/abs/1311.1238}{{\tt arXiv:1311.1238 [gr-qc]}}.
%%CITATION = ARXIV:1311.1238;%%.

\bibitem{Hawking:1981gb}
S.~W. Hawking, ``{The Boundary Conditions of the Universe},''
{\em Pontif. Acad. Sci. Scr. Varia} {\bf 48} (1982)  563--574.
%%CITATION = PRINT-82-0179 (CAMBRIDGE);%%.

\bibitem{Hartle:1983ai}
J.~B. Hartle and S.~W. Hawking, ``{Wave Function of the Universe},''
\href{http://dx.doi.org/10.1103/PhysRevD.28.2960}{{\em Phys. Rev.} {\bf D28}
  (1983)  2960--2975}.
%%CITATION = PHRVA,D28,2960;%%.

\bibitem{Vilenkin:1982de}
A.~Vilenkin, ``{Creation of Universes from Nothing},''
\href{http://dx.doi.org/10.1016/0370-2693(82)90866-8}{{\em Phys. Lett.} {\bf
  117B} (1982)  25--28}.
%%CITATION = PHLTA,117B,25;%%.

\bibitem{Gibbons:1977zz}
G.~W. Gibbons, ``{The Einstein Action of Riemannian Metrics and Its Relation to
  Quantum Gravity and Thermodynamics},''
\href{http://dx.doi.org/10.1016/0375-9601(77)90244-4}{{\em Phys. Lett.} {\bf
  A61} (1977)  3--5}.
%%CITATION = PHLTA,A61,3;%%.

\bibitem{Feldbrugge:2017kzv}
J.~Feldbrugge, J.-L. Lehners, and N.~Turok, ``{Lorentzian Quantum Cosmology},''
  \href{http://dx.doi.org/10.1103/PhysRevD.95.103508}{{\em Phys. Rev.} {\bf
  D95} (2017) no.~10, 103508},
\href{http://arxiv.org/abs/1703.02076}{{\tt arXiv:1703.02076 [hep-th]}}.
%%CITATION = ARXIV:1703.02076;%%.

\bibitem{Feldbrugge:2018gin}
J.~Feldbrugge, J.-L. Lehners, and N.~Turok, ``{Inconsistencies of the New
  No-Boundary Proposal},''
\href{http://arxiv.org/abs/1805.01609}{{\tt arXiv:1805.01609 [hep-th]}}.
%%CITATION = ARXIV:1805.01609;%%.

\bibitem{Feldbrugge:2017fcc}
J.~Feldbrugge, J.-L. Lehners, and N.~Turok, ``{No smooth beginning for
  spacetime},'' \href{http://dx.doi.org/10.1103/PhysRevLett.119.171301}{{\em
  Phys. Rev. Lett.} {\bf 119} (2017) no.~17, 171301},
\href{http://arxiv.org/abs/1705.00192}{{\tt arXiv:1705.00192 [hep-th]}}.
%%CITATION = ARXIV:1705.00192;%%.

\bibitem{DiazDorronsoro:2017hti}
J.~Diaz~Dorronsoro, J.~J. Halliwell, J.~B. Hartle, T.~Hertog, and O.~Janssen,
  ``{Real no-boundary wave function in Lorentzian quantum cosmology},''
  \href{http://dx.doi.org/10.1103/PhysRevD.96.043505}{{\em Phys. Rev.} {\bf
  D96} (2017) no.~4, 043505},
\href{http://arxiv.org/abs/1705.05340}{{\tt arXiv:1705.05340 [gr-qc]}}.
%%CITATION = ARXIV:1705.05340;%%.

\bibitem{DiazDorronsoro:2018wro}
J.~Diaz~Dorronsoro, J.~J. Halliwell, J.~B. Hartle, T.~Hertog, O.~Janssen, and
  Y.~Vreys, ``{Damped perturbations in the no-boundary state},''
\href{http://arxiv.org/abs/1804.01102}{{\tt arXiv:1804.01102 [gr-qc]}}.
%%CITATION = ARXIV:1804.01102;%%.

\bibitem{Feldbrugge:2017mbc}
J.~Feldbrugge, J.-L. Lehners, and N.~Turok, ``{No rescue for the no boundary
  proposal: Pointers to the future of quantum cosmology},''
  \href{http://dx.doi.org/10.1103/PhysRevD.97.023509}{{\em Phys. Rev.} {\bf
  D97} (2018) no.~2, 023509},
\href{http://arxiv.org/abs/1708.05104}{{\tt arXiv:1708.05104 [hep-th]}}.
%%CITATION = ARXIV:1708.05104;%%.

\bibitem{Halliwell:1989vu}
J.~J. Halliwell and J.~Louko, ``{Steepest Descent Contours in the Path Integral
  Approach to Quantum Cosmology. 2. Microsuperspace},''
\href{http://dx.doi.org/10.1103/PhysRevD.40.1868}{{\em Phys. Rev.} {\bf D40}
  (1989)  1868}.
%%CITATION = PHRVA,D40,1868;%%.

\bibitem{Teitelboim:1981ua}
C.~Teitelboim, ``{Quantum Mechanics of the Gravitational Field},''
\href{http://dx.doi.org/10.1103/PhysRevD.25.3159}{{\em Phys. Rev.} {\bf D25}
  (1982)  3159}.
%%CITATION = PHRVA,D25,3159;%%.

\bibitem{Teitelboim:1983fk}
C.~Teitelboim, ``{The Proper Time Gauge in Quantum Theory of Gravitation},''
\href{http://dx.doi.org/10.1103/PhysRevD.28.297}{{\em Phys. Rev.} {\bf D28}
  (1983)  297}.
%%CITATION = PHRVA,D28,297;%%.

\end{thebibliography}\endgroup

\end{document}